# Ab initio based analysis of grain boundary segregation in Al-Mg and Al-Zn binary alloys

M. V. Petrik[1,2], A. R. Kuznetsov[1,2,3], N. Enikeev[4,5], Yu. N. Gornostyrev[1,2,3], R. Z. Valiev[4,5]

[1]Institute of Metal Physics, Ural Branch, Russian Academy of Sciences, Ekaterinburg 620041, Russia.

[2]Institute of Quantum Materials Science, Ekaterinburg 620075, Russia.

[3]Ural Federal University, Ekaterinburg 620002, Russia

[4]Institute for Physics of Advanced Materials, Ufa State Aviation Technical University, Ufa 450000, Russia.

[5]Saint Petersburg State University, 198504 Peterhof, Saint Petersburg, Russia

Based on ab-initio simulations, we report on the nature of principally different mechanisms for interaction of Mg and Zn atoms with grain boundaries in Al alloys leading to different morphology of segregation. The Mg atoms segregate in relatively wide GB region with heterogeneous agglomerations due to the deformation mechanism of solute-GB interaction. In contrast, in the case of Zn atoms an electronic mechanism associated with the formation of directional bonding is dominating in the solute-GB interaction. As a result, for Zn atoms it is energetically beneficial to occupy interstitial positions at the very GB and to be arranged into thin layers along the GBs. The results obtained show the essential role of elements chemistry in segregation formation and explain the qualitative features in morphology of GB segregation observed in Al-Mg and Al-Zn alloys with ultrafine grains.

## I. INTRODUCTION

Grain boundary (GB) segregation of solute atoms is important phenomenon that affects many physical properties of polycrystalline materials [1]. The formation of the segregations on GBs can change significantly strength and plasticity, conditions of the phase equilibrium [2] of the alloy and thermal stability of structure [3,4,5,6]. The role of GB segregation rises especially in ultrafine grained materials (UFG) [7,8,9,10] where fractions of GB and bulk atoms are comparable. As result, UFG materials are interface-controlled and the fundamental understanding of the GB segregation is necessary for effective control of structural state and properties of alloys for advanced applications.

The investigation of GB chemistry is a challenging task for both experiment and theoretical modeling. Although the problem of GB segregation has been in the focus of research studies for many years, the physical mechanisms of this phenomena remain the subject of debates (see Refs. [1,10,11,12,13]). Over the last years new important information about the GB segregation has been obtained by using 3D atom probe tomography [14,15]. The recent advanced studies [16,17,18,19,20] focused on the investigations of UFG alloys produced by severe plastic deformation (SPD) where GB segregations are more pronounced in comparison with coarse-grained materials. It has been shown that Al-Mg and Al-Zn alloys give bright example of dramatic dependence of segregations and their effect on mechanical properties of the chemistry of alloying elements. In particular, for an Al–Mg system it was observed by using scanning transmission electron microscopy and atom probe tomography that Mg atoms tend to form heterogeneous agglomerations at GBs [7,21,22] and it results in extra-hardening. In contrast, thin layers of Zn atoms are distributed homogeneously along the GBs in the UFG Al-Zn alloy produced by SPD [10,5]. These specific segregations are thought to be a reason for GB sliding enhancement in UFG Al-Zn at relatively lower temperatures alloys and resulted in unusual super-ductility [23,24,25].

Striking differences in the segregation morphology and segregation-controlled GB phenomena in Al-Mg and Al-Zn alloys could hardly be explained with traditional approaches used to describe GB segregation processes [1,26]. A consistent analysis of the GB segregation formation requires information about the energy change of solute atoms near the GB for different solute chemistry and GB types. This information can be well characterized by using of first principles calculations or atomistic molecular dynamics (MD) simulations. MD approach allow to implement large scale modeling; however, it is limited by the reliability of interatomic interaction potentials. This problem becomes most essential when considering interaction of atoms with GBs because the validity of conventional interatomic potentials for the bulk material is questionable near GBs.

First principles density functional theory (DFT) methods is now reliable tools for investigations of structure and energetics of real crystals. In particular, these approaches are widely used to study the solute–GB interactions, the effects of alloying elements on the electronic structure and cohesion of GBs [11,12,13, 27,28,29]. As it was discussed in [11,13], a realistic picture of solute–GB interactions can be rather complex since both *deformation* (related to *size effect*) and *electronic* (associated with a charge transfer or changes in the *chemical bonding*) mechanisms can play an important role.

Present study aims to reveal the specific features of segregation processes in Al alloys by using the first principles simulations of the GB atomic structure occupied by two types of alloying elements (Mg and Zn) and to find out the physical explanation for the observed diversity in the morphology of GB segregation. Here we consider equilibrium segregation of solute atoms to a GB. In this case the segregation isotherm [1,24] relates changes in the solute concentration at the GB with changes in the Gibbs free energy. It is usually assumed that segregation of individual atoms is driven by energy gain upon moving a solute atom from the bulk to the GB. Wherein all entropy contributions (such as anharmonic, and vibrational) can be neglected, except the ideal configurational entropy, which is then used for calculating the enrichment of the GB phase by the solute.

The direct ab initio investigation of the chemistry and structure of real strain-distorted GBs produced by SPD is not feasible since the size of required crystallites is too large in this case. Nevertheless, we can figure out important trends of solute–GB interactions by considering a special GB (see discussions in Ref. [13]). Though the interaction region is much shorter in comparison to the distorted GB, main features of solute–GB interactions such as local deformations and changes in chemical bonding will be similar for special and distorted GBs.

The modeling of GB with solute atom was performed by density functional theory (DFT) method as implemented in the pseudopotential code SIESTA [30]. All calculations were done by using the generalized gradient approximation (GGA-PBE) [31]. The ion cores were described by norm-conserving pseudo-potentials [32] and the wave functions are expanded with a double-ζ plus polarization basis of localized orbitals for aluminum, zinc, and magnesium. Optimization of the forces and total energy was performed with an accuracy of 0.04 eV/Å and $10^{-5}$ eV, respectively. All calculations were carried out with an energy mesh cut-off of 600 Ry and k-point mesh of 6×6×6 in the Monkhorst-Pack scheme [33].

To reveal the features of solute–GB interactions in Al we calculated the electronic structure and total energy of the crystallite (with structural relaxation) containing special Σ5{210}[001] tilt GB and one or two solute atoms of Mg or Zn, which replaced the Al atom in one of the sites 1–10 near and in the GB center (see insert in Fig. 1a). We used 80- and 160-atom supercell with the dimensions 8.94a x 2.23a x 1.0a and 8.94a x 2.23a x 2.0a with periodic boundary conditions (*a* is the lattice parameter of Al). The supercell contained two crystallites misoriented with respect to each other by $53.1^0$ about the [001] axis. In the center of GB, we considered both substitutional and interstitial solute position. The segregation energy $\Delta E_s(R_n)$ was determined as the total energy difference between the crystallite containing the impurity at the distance $R_n$ from the GB plane (corresponding position *n*) and the crystallite with the same impurity situated on site 11 which matches to the maximum distance from the GB plane:

$$\Delta E_s(R_n) = E_{GB}^s(R_n) - E_{bulk}^s, \qquad (1)$$

Calculated segregation energies $\Delta E_s(R)$ for Mg and Zn atoms are presented in Fig. 1a as a function of the distance $R_n$ from the GB plane. It can be seen that the interaction between each of the substitutional atoms and the Σ5{210}[001] tilt GB is short-ranged (concentrated in a narrow layer near the GB) and vary non-monotonically with the distance. Oscillations of the segregation energy $\Delta E_s(R)$ are frequently observed for a special GB (see for example Ref. [13]) and may be caused by the heterogeneity of local distortions or by so-called Friedel oscillations of the electron density near the GB [26]. We found that Mg atoms have the strongest energy preference when they substitute Al in the center of the GB (position 1) in contrast to Zn which segregation energy $\Delta E_s$ in the position 1 is positive [34].

Table 1. DFT-calculated solution energy $E_{sol}$, and relative atomic displacements in the first ($\varepsilon_1$) and the second ($\varepsilon_2$) coordination spheres around the substitutional impurities in the bulk of Al. $R_{at}$ is an empirical atomic radius.

| solute | el. conf | $R_{at}$, pm | $E_{sol}$, eV | $\varepsilon_1$, % | $\varepsilon_2$, % |
|---|---|---|---|---|---|
| Mg | $3s^2$ | 150 | -0.03 | 1.4 | 0.3 |
| Zn | $3d^{10}\,4s^2$ | 135 | 0.02 | 0.0 | 0.0 |

As was shown in Ref. [13], the *deformation mechanism* is dominating in the interaction between GB and Mg solute. Since Mg atom in Al behaves as a larger one, its movement to the GB results in energy decreasing due to the larger free volume available at the segregation site. Zn solute shows an opposite trend; the substitution of Al by Zn in bulk does not practically change positions of the nearest and the next nearest neighbors in fcc Al (see Table 1), i.e., the relaxation energy is very small. By considering the distribution of charge density of valence electrons near the GB we found that valence electrons density in the GB center only slightly decreases due to the Mg atom. At the same time, the electronic mechanism of solute-GB interaction is dominating for Zn solute. As a result, the energy gain is to be associated with the formation of certain Al–Zn quasi-covalent bonds due to 3$p$Al–3$d$Zn hybridization of electronic states and it enhance due to local changes in coordination and distances from Zn to the nearest neighbors of host atoms.

It was found that Zn solute can pass from substitutional position 1 or 2 to interstitial position 0 that accompanied with formation of a vacancy (V) and the following atomic relaxation of the GB structure lead to the energy gain (see Fig. 1a). The stability of this configuration is achieved due to the formation of quasi-covalent bonds between Zn and nearest Al atoms. In contrast to Zn, the interstitial position appears to be not stable for Mg. Thus, Zn gives an example of the element, which segregates at the GB in the interstitial position that can modify the GB structure.

The features of distribution of alloying elements near GB are determined not only by solute–GB interactions but also by the interaction of solutes with each other; the latter contribution limits the concentration of the segregating component if its interaction energy is positive (see discussion in Ref. [13]) and enhances it in the opposite case. To figure out the segregation ability of Mg and Zn on the GB in Al we calculated interaction energy between two solutes (i) in the bulk and (ii) with one of them located at the GB (substitutional position in case of Mg and interstitial one in case of Zn).

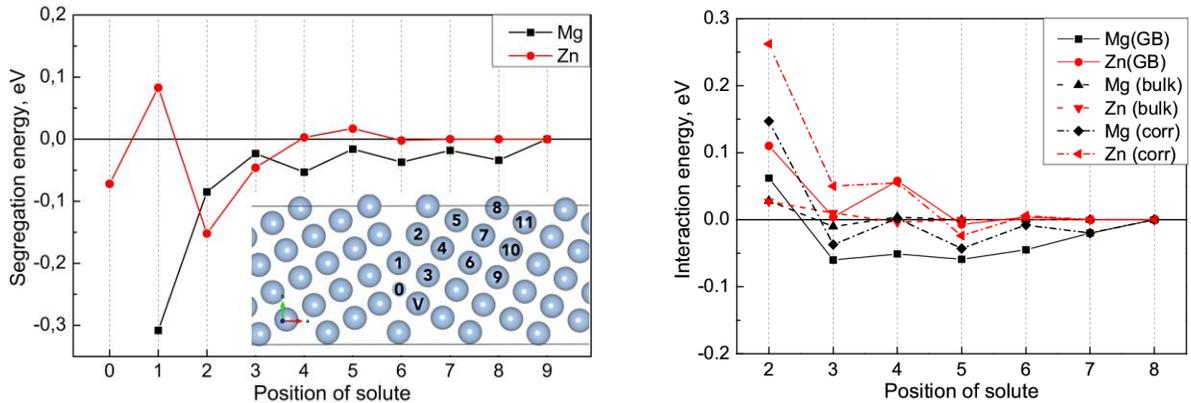

**Fig. 1.** Segregation energies $\Delta E_s$ of solutes with the $\Sigma5\{210\}[001]$ tilt GB **(a)** and effective interaction energy between two solute atoms **(b)** calculated by SIESTA method for 80-atom crystalline. The sequence of symbols with distance corresponds to the substitutional positions 1–9; 0 corresponds to the interstitial position and 'V' stands to vacation (see insertion). Solid lines in Fig. 2b correspond to the interaction energy between two solutes in dependence on distance when one of them is located in the GB center. Dashed lines in Fig. b are obtained from solid ones by subtraction of corresponding segregation energies (a) and describe true solute-solute interaction near GB. Dash dotted lines show interaction energies between solutes in a bulk.

As seen from results of calculations (Fig. 1b), in case (ii) Mg–Mg interaction energy changes essentially near the GB and becomes negative when the distance between solutes corresponds to the $2^{nd}$–$5^{th}$ nearest neighbors. As a result, we should expect some enhancement of segregation ability of Mg in Al due to effect of the GB on the Mg–Mg interaction in its vicinity. In case of Zn, the interaction energy keeps a positive value and not contribute to the segregation enhancement.

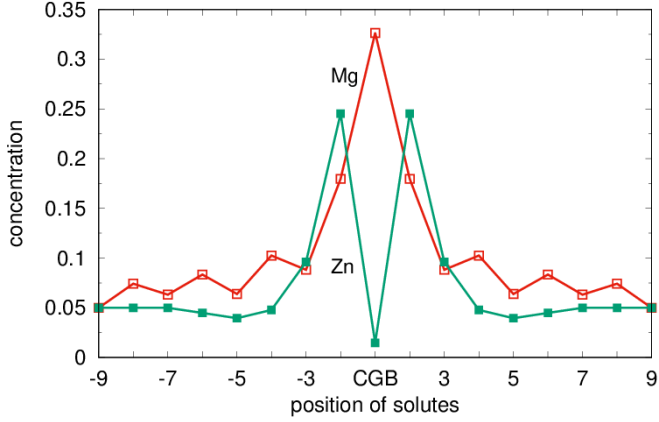

**Fig. 2.** Distribution of solutes near the $\Sigma 5\{210\}[001]$ tilt GB calculated by using the Fowler model [1] at $C=0.05$, $T=700$ K. CGB labeled the center GB

To illustrate the feature of the segregation morphology of considered elements on GB in Al we used simple model approach and calculated segregation and solute-solute interaction energies (Fig. 2). In the framework of the regular solid solution model, GB segregation can be described by the Fowler equation [1]

$$\frac{C_{GB}}{1-C_{GB}} = \frac{C}{1-C} \exp\left[\frac{-\Delta E_s + V(C - C_{GB})}{kT}\right], \quad (2)$$

where $V = \sum_n Z_n V_n$ is the mixing energy, $V_n$ is the effective interaction energy between the solute atoms, $Z_n$ is the coordination number for $n$-th coordination shell, $C_{GB} = C_{GB}(R_n)$ is concentration of solute atoms near GB and $C$ is mean concentration. The distribution of Mg and Zn near center of considered GB at 700 K calculated using Eq. (2) is shown in Fig. 2. One can see that segregation profiles are rather different for considered solutes. We found relatively wide segregation in case of Mg (the width of the enriched region is about 1.5 nm), which profile is qualitatively similar to the previously obtained results [13]. At the same time, the tiny segregations which mostly concentrated in two layers were found in case of Zn.

Thus, by using ab-initio simulation we revealed that the origin of experimentally observed [4,7,10,24] drastically different segregation behavior of Mg and Zn in Al alloys is related to the fundamental features of electronic structure of solutes and chemical bonding rather than to special details of the GB structure. In case of Mg, which behaves as a large atom in the Al lattice, the energy gain due to atomic relaxation at the GB determines its segregation ability (deformation interaction mechanism). Furthermore, lattice distortions near GB results in long-range Mg–GB interaction (see Fig. 1a) which promoting the formation of extended segregations. At the same time, Zn solutes show an opposite behavior; the atomic relaxation is neglected and electronic mechanism related to hybridization of $3p$–$3d$ electronic states gives the main contribution into the "solute–GB" interaction. In this case the tendency to form Zn–Al quasi-covalent bonds makes the positions with higher local atomic coordination more preferable. As result, small number of sites available for segregation is the general feature of GB in Al-Zn alloys responsible for appearing of thin GB segregations of Zn atoms.

As a result, Mg atoms tend to form clouds/agglomerations in the area adjacent to the boundary due to strong Mg–GB and Mg–Mg interactions. In contrast, Zn atoms prefer to take positions in the very grain boundary, repulsive Zn–Zn interaction near the GB does not allow keeping additional Zn atoms in the Al matrix around, and segregation takes place within the few crystal layers near the interface.

The research was carried out within the state assignment of FASO of Russia (theme "Structure" No. 01201463331). The results have been obtained using the computational resources of MCC National Research Center "Kurchatov Institute" (http://computing.kiae.ru ) and Uran supercomputer of Institute of mathematics and mechanics UB RAS. This work was supported through the Contract No 14.B25.31.0017 of the Russian Ministry for Education and Science dated by 28.06.2013.